\documentclass[12pt]{article}
\begin{document}
\begin{center}
\large\bf{The Effective Potential in the \\
Massive $\phi_4^4$ Model}
\end{center}
\vspace{2cm}
\begin{center}
F. T. Brandt\\
Instituto de Fisica\\
Universidade de S$\rm{\widetilde{a}}$o Paulo\\
S$\rm{\widetilde{a}}$o Paulo, 05389-970 SP Brazil\\
\vspace{1cm}
F. A. Chishtie\\
D.G.C. McKeon\\
Department of Applied Mathematics\\
University of Western Ontario\\
London  N6A 5B7  Canada
\end{center}
Tel: 519-661-2111, ext. 88789 \hfill  PACS No. \\
Fax: 519-661-3523\\
Emails: fbrandt@satie.if.usp.br, fachisht@uwo.ca, dgmckeo2@uwo.ca
\vspace{2cm}
\section{Abstract}
 By applying the renormalization group equation, it has been shown that the effective potential $V$ in the massless $\phi_4^4$ model and in massless scalar quantum electrodynamics is independent of the scalar field.  This analysis is extended here to the massive $\phi_4^4$ model, showing that the effective potential  is independent of $\phi$ here as well.
\eject

Recently, it has been shown [1] that in the massless scalar model with classical Lagrangian 
$$\textsl{L} = \frac{1}{2} (\partial_\mu\phi )^2 - \frac{\lambda}{4!} \phi^4\eqno(1)$$
and in massless scalar quantum electrodynamics (MSQED) with classical Lagrangian
$$\textsl{L} = \left(\partial_\mu + ieA_\mu\right)\phi^* \left(\partial^\mu - ieA^\mu\right)\phi - \frac{1}{4} \left(\partial_\mu A_\nu - \partial_\nu A_\mu \right)^2 - \lambda (\phi^* \phi)^2\eqno(2)$$
the effective potential $V(\phi)$ [2-5] is either flat (ie, independent of the background field $\phi$) with non-analytic dependence on the coupling, or there is no spontaneous symmetry breaking (ie, $< \phi > \equiv  v = 0$). This has been done by applying the renormalization group (RG) to $V$ and requiring that $V^\prime (\phi = v) = 0$.

In this note, we supplement the Lagrangian of eq. (1) with a mass term
$$\textsl{L} = -\frac{1}{2} m^2\phi^2\eqno(3)$$
and apply the same techniques as were used in ref. [1] to see the effect of inclusion of this mass term on the effective potential.

This same action has been considered in refs. [6-11]. It was demonstrated in ref. [6] that the general form of the resulting effective potential is
$$V(\lambda, x, y,\phi ) = \left\lbrace \left(a + \frac{b}{x}\right) + 
\sum_{\ell = 1}^\infty \lambda^{\ell + 1} \sum_{m =0}^{\ell - 1} x^{m-2} 
\sum_{n=0}^\ell y^n a_{\ell mn}\right.\eqno(4)$$
$$\left. + \left(\frac{1-x}{x}\right)^2 \sum_{k= 1}^\infty
 b_k\lambda^k\right\rbrace \phi^4\nonumber$$
where $a = \frac{-5}{24}$, $b = \frac{1}{4}$ and
$$x = \frac{1}{1 + \frac{2m^2}{\lambda\phi^2}}\;\;\;\; y = \ln \left(\frac{\lambda\phi^2}{2\mu^2} \frac{1}{x} \right) = \ln \left(\frac{m^2 + \frac{\lambda\phi^2}{2}}{\mu^2}\right).
\eqno(5)$$
Following refs. [1,12], we now expand $V$ in powers of $y^n$, so that
$$V(\lambda , x,y,\mu ,\phi ) = \left( \sum_{n = 0}^\infty
A_n (\lambda , x)y^n\right)\phi^4 .\eqno(6)$$
The RG equation for $V$ is
$$\mu^2 \frac{dV}{d\mu^2} = \left[\mu^2 \frac{\partial}{\partial \mu^2} + \beta(\lambda) \frac{\partial}{\partial\lambda} + \gamma_m (\lambda) m^2 \frac{\partial}{\partial m^2} + \gamma_\phi (\lambda)\phi^2 \frac{\partial}{\partial\phi^2}\right] V = 0\eqno(7)$$
or by eqs. (5) and (6)
$$\left\lbrace\left[ -1+\left(\frac{\beta}{\lambda} + \gamma_\phi\right) x + \gamma_m (1-x)\right] \frac{\partial}{\partial y} + \left[\frac{\beta}{\lambda} - \gamma_m + \gamma_\phi\right] x(1-x)\frac{\partial}{\partial x}\right.\nonumber$$
$$\left. + \beta \frac{\partial}{\partial\lambda} + 2\gamma_\phi\right\rbrace 
\sum_{n=0}^\infty A_n(\lambda , x)y^n = 0.\eqno(8)$$

Eq. (8) is satisfied order by order in $y$ provided 
$$A_{n+1}(\lambda , x) = \frac{1}{n+1} \left[
f(\lambda , x) \frac{\partial}{\partial x} + g(\lambda , x) \frac{\partial}{\partial \lambda} + h(\lambda , x)\right]A_n(\lambda, x)\eqno(9)$$
where
$$f(\lambda , x) = \frac{\left(\frac{\beta}{\lambda} - \gamma_m + \gamma_\phi\right)x(1-x)}
{1 - \left(\frac{\beta}{\lambda} + \gamma_\phi\right)x-\gamma_m(1-x)}\eqno(10)$$
$$g(\lambda , x) = \frac{\beta}{1 - \left(\frac{\beta}{\lambda} + \gamma_\phi\right)x-\gamma_m(1-x)}\eqno(11)$$
$$h(\lambda , x) = \frac{2\gamma_\phi}{1 - \left(\frac{\beta}{\lambda} + \gamma_\phi\right)x-\gamma_m(1-x)} .\eqno(12)$$

Furthermore, since
$$\frac{dV}{d\phi^2} = \left\lbrace \frac{1}{\phi^2}\left[ x (1-x) \frac{\partial}{\partial x} + x \frac{\partial}{\partial y}\right] + \frac{\partial}{\partial\phi^2}\right\rbrace \left[\sum_{n=0}^\infty A_n(\lambda , x)y^n\right]\phi^4\nonumber$$
$$= \sum_{n=0}^\infty \left[\left(
x(1-x)\frac{\partial A_n}{\partial x} + 2 A_n\right)y^n + n x A_n y^{n-1}\right]\phi^2 ,\eqno(13)$$
an extremum of $V$ is attained at $\phi^2 = v^2$ if at each order of $y$, eq. (13) vanishes, leading to
$$A_{n+1} (\lambda , x_0) = \frac{1}{n+1} \left[ - \frac{2}{x} - (1-x)\frac{\partial}{\partial x}\right]A_n (\lambda , x_0)\eqno(14)$$
provided $v^2 \neq 0$. (We have set $x_0 = \frac{1}{1 + \frac{2m^2}{\lambda v^2}}$.)  Having both the renormalization group equation (8) and the extremum condition (13) hold order-by-order in $y$ thus leads to a pair of distinct equations relating $A_{n+1}$ to $A_n$.

If eq. (14) were to hold for all $x$, then
$$A_n(\lambda ,x) = \frac{1}{n!} \left(- \frac{2}{x} - (1-x)\frac{\partial}{\partial x}\right)^n A_0 (\lambda , x)\eqno(15)$$
so that eq. (6) reduces to
$$V = \sum_{n=0}^\infty \frac{y^n}{n!} \left( - \frac{2}{x} - (1-x)\frac{\partial}{\partial x}\right)^n A_0 (\lambda , x)\phi^4 .\eqno(16)$$
If now $z = \ln (1-x)$, then eq. (16) becomes
$$V = \exp \left( 2 \int_{z_0}^z \frac{dt}{1-e^t}\right)\sum_{n=0}^\infty \frac{y^n}{n!} \left(\frac{\partial}{\partial z}\right)^n B_0 (\lambda , z)\phi^4 =
\exp\left(-2 \int_z^{z+y} \,\frac{dt}{1-e^t}\right)
A_0 (\lambda , z + y)\phi^4
\eqno(17)$$
where
$$B_0(\lambda , z) = \exp \left( -2 \int_{z_0}^z \frac{dt}{1 - e^t}\right)A_0(\lambda , x).
\eqno(18)$$
Noting that from eq. (5),
$$z + y = \ln \left(1 - \frac{1}{1 + \frac{2m^2}{\lambda\phi^2}}\right) + \ln\left(\frac{m^2 + \frac{\lambda\phi^2}{2}}{\mu^2}\right) = \ln \left(\frac{m^2}{\mu^2}\right)\eqno(19)$$
eq. (17) becomes
$$V = \frac{4(m^2 - \mu^2)^2}{\lambda^2}\,A_0\left(\lambda , \ln \frac{m^2}{\mu^2}\right)\eqno(20)$$
showing that $V$ has no dependence on $\phi$.  The function $A_0$ can be fixed by eqs. (9) and (14) with $n = 0$; this leads to the equation
$$\left( - \frac{2}{x} - (1-x) \frac{\partial}{\partial x}\right) A_0 (\lambda , x) = \left( f(\lambda , x)\frac{\partial}{\partial x} + g(\lambda , x)\frac{\partial}{\partial \lambda} + h (\lambda , x)\right)A_0 (\lambda , x).
\eqno(21)$$

An alternative approach to deriving eq. (20)
which does not rely on eq. (13) holding order by order in $y$ at $\phi^2 = v^2$ can also be used, following a line of reasoning based on the ``method of characteristic"' [13,14] much as in ref. [1]. We begin by choosing $\mu^2$ so that $y$ in eq. (13) vanishes when $\phi^2 = v^2$; eq. (14) then holds when $n = 0$, provided $\lambda$ and $x_0$ are evaluated with $\lambda (\mu^2)$ and $m^2(\mu^2)$ evaluated at this value of $\mu^2$. In refs. [2,15] this is taken to impose a restriction on the couplings at this value of $\mu^2$; we instead follow ref. [1] and take this to be an equation that fixes the function $A_1$ in terms of $A_0$, since the actual values of $\lambda(\mu^2)$ and $m^2(\mu^2)$ can be independently varied at this value of $\mu^2$ by adjusting the initial conditions on these running parameters. We thus have a functional equation relating $A_0(\lambda , x)$ to $A_1 (\lambda , x)$ rather than equation that relates the values of parameters in the theory,
$$A_1 (\lambda , x) = \left[ - \frac{2}{x} - (1-x) \frac{\partial}{\partial x}\right]A_0 (\lambda , x),\eqno(22)$$
in addition to eq. (9).

Much as was done in the discussion of MSQED in ref. [1], we now define
$$a_n(\overline{\lambda}(t),\overline{x}(t), t) =
\exp \left[
\int_{t_{0}}^t d\tau \,h(\overline{\lambda}(\tau), \overline{x}(\tau))\right]A_n 
(\overline{\lambda}(t), \overline{x}(t)),\eqno(23)$$
where
$$\frac{d\overline{x}(t)}{dt} = f(\overline{\lambda}(t),\overline{x}(t))\;\;\;\;
(\overline{x}(t_0) = x)\eqno(24)$$
and
$$\frac{d\overline{\lambda}(t)}{dt} = g(\overline{\lambda}(t),\overline{x}(t))\;\;\;\;
(\overline{\lambda}(t_0) = \lambda).\eqno(25)$$
Together, eqs. (9) and (23-25) show that
$$\frac{d}{dt} a_n(\overline{\lambda}(t), \overline{x}(t), t)
= (n+1)a_{n+1}(\overline{\lambda}(t),\overline{x}(t),t)
\eqno(26)$$
with
$$a_n(\overline{\lambda}(t_0),\overline{x}(t_0),t_0) = A_n(\lambda ,x).\eqno(27)$$
If now
$$\overline{y}(t) = \ln\left( \frac{\overline{\lambda}(t)\overline{\phi}^2(t)}
{2\overline{\mu}^2(t)}\,\frac{1}{\overline{x}(t)}\right)\eqno(28)$$
with
$$\frac{1}{\overline{\phi}^2(t)} \frac{d\overline{\phi}^2(t)}{dt} = 
\frac{\gamma_\phi(\overline{\lambda}(t)
)}
{\left[1-\left( \frac{\beta(\overline{\lambda}(t))}{\overline{\lambda}(t)} + \gamma_\phi(\overline{\lambda}(t))\right) \overline{x}(t)-
\gamma_m(\overline{\lambda}(t))(1 - \overline{x} (t))\right]}\eqno(29)$$
and 
$$\frac{1}{\overline{\mu}^2(t)} \frac{d\overline{\mu}^2(t)}{dt} = 
\frac{1}
{\left[1-\left( \frac{\beta(\overline{\lambda}(t))}{\overline{\lambda}(t)} + \gamma_\phi(\overline{\lambda}(t))\right) \overline{x}(t)-
\gamma_m(\overline{\lambda}(t))(1 - \overline{x} (t))\right]}\eqno(30)$$
$$\left(\overline{\phi}^2 (t_0) = \phi^2,\;\;\;\;\overline{\mu}^2 (t_0) = \mu^2\right)\nonumber$$
then eqs. (24, 25, 28-30) together imply that
$$\frac{d\overline{y}(t)}{dt} = -1\;\;\;\;\;(\overline{y} (t_0) = y).\eqno(31)$$
Now defining
$$V(t) = \sum_{n=0}^\infty a_n(\overline{\lambda}(t), \overline{x}(t), t)\overline{y}^n(t)\phi^4,\eqno(32)$$
then by eqs. (26) and (31), we have
$$\frac{dV(t)}{dt} = 0\eqno(33)$$
with
$$V(t_0) = V(\lambda , x, y, \mu ,\phi).\eqno(34)$$
From eqs. (26) and (32), it follows that
$$V(t) = \sum_{n=0}^\infty \frac{\overline{y}^n(t)}{n!}
\left(\frac{d}{dt}\right)^n a_0(\overline{\lambda}(t), \overline{x}(t), t)\phi^4\eqno(35)$$
$$= a_0(\overline{\lambda}(t + \overline{y}(t)), \overline{x}(t + \overline{y} (t)), t + \overline{y}(t))\phi^4\eqno(36)$$
which by eq. (33) and (34) leads to
$$V(\lambda , x, y, \mu , \phi) = a_0(\overline{\lambda}(t_0 +y), \overline{x}(t_0+y), t_0 + y)\phi^4 .\eqno(37)$$

Again following the steps used to analyze MSQED in ref. [1], we define
$$\widetilde{A}_0 (\overline{\lambda}(t), \overline{x}(t), t) = \left[
\exp \int_{t_{0}}^t \frac{2d\tau}{\overline{x}(\tau)}\right]a_0(\overline{\lambda}(t), \overline{x}(t), t)\eqno(38)$$
which by eq. (23) is also equal to
$$= \left[\exp \int_{t_{0}}^t \left( \frac{2}{\overline{x}(\tau)} + h(\overline{\lambda}(\tau), \overline{x}(\tau), \tau))\right)d\tau\right] A_0 
(\overline{\lambda}(t),\overline{x}(t)). \eqno(39)$$
Eqs. (21), (24), (25) and (39) now show that
$$\frac{d}{dt}\widetilde{A}_0 (\overline{\lambda}(t), \overline{x}(t), t) = 
- (1 - \overline{x} (t))
\frac{\partial}{\partial\overline{x}(t)}
\widetilde{A}_0 (\overline{\lambda}(t), \overline{x}(t), t).\eqno(40)$$
It is also apparent from eqs. (27) and (38) that
$$\widetilde{A}_0 (\overline{\lambda}(t_0), \overline{x}(t_0), t_0) = 
a_0 (\overline{\lambda}(t_0), \overline{x}(t_0), t_0)= A_0(\lambda , x).
\eqno(41)$$

We now can use eqs. (35) and (38) to write
$$V(t) = \sum_{n=0}^\infty \frac{\overline{y}^n(t)}{n!} \left(\frac{d}{dt}\right)^n\left\lbrace \left[ \exp - \int_{t_{0}}^t \frac{2d\tau}{\overline{x}(\tau)}\right]
\widetilde{A}_0 (\overline{\lambda}(t), \overline{x}(t), t)\right\rbrace\phi^4\eqno(42)$$
which by eq. (40) becomes
$$= \left[ \exp - \int_{t_{0}}^t \frac{2d\tau}{\overline{x}(\tau)}\right]\sum_{n=0}^\infty
\frac{\overline{y}^n(t)}{n!}\left[\frac{-2}{\overline{x}(t)}- (1-\overline{x}(t))\frac{\partial}{\partial\overline{x}(t)}\right]^n
\widetilde{A}_0(\overline{\lambda}(t), \overline{x}(t), t)\phi^4.\eqno(43)$$
Now following the same argument that led from eqs. (16) to (20), eq. (43) reduces to
$$V(t) = \left[ \exp - \int_{t_{0}}^t \frac{2d\tau}{\overline{x}(\tau)}\right]
\exp\left(-2 \int_{\overline{z}(t)}^{\overline{z}(t)+\overline{y}(t)}\frac{d\tau}{1-e^\tau}\right)
\widetilde{A}_0(\overline{\lambda}(t), \overline{z}(t) + \overline{y}(t), t)\phi^4.\eqno(44)$$
where $\overline{z}(t) = \ln(1-\overline{\lambda}(t))$. We can now set $t = t_0$ in eq. (44); by eqs. (19), (34) and (41) we recover eq. (20).

It is of interest to consider the solution of the $m^2 \rightarrow 0$ limit of eq. (7),
$$\left[\mu^2 \frac{\partial}{\partial \mu^2} + \beta (\lambda) \frac{\partial}{\partial \lambda} + \gamma_\phi (\lambda) \phi^2 \frac{\partial}{\partial\phi^2}\right]\left[S\left(\lambda , \ln \left( \frac{\lambda\phi^2}{2\mu^2}\right)\right)\phi^4\right] = 0\eqno(45)$$
in the case where $\beta(\lambda)$ and $\gamma_\phi(\lambda)$ are taken to be given exactly by the order $\lambda^2$ expressions $\beta(\lambda) = b\lambda^2$, $\gamma_\phi (\lambda) = g\lambda^2$. We first note that eq. (45) becomes
$$\left[\left( \frac{-1+\gamma+\beta/\lambda}{\beta}\right)\frac{\partial}{\partial L} + \frac{\partial}{\partial \lambda} + \left( \frac{2\gamma}{\beta}\right)\right] S(\lambda , L) = 0.\eqno(46)$$
$$\;\;\;\;\;\;\;\;\;\left(L = \ln \frac{\lambda \phi^2}{2\mu^2}\right)\nonumber$$
If now
$$S(\lambda , L) = \exp\left(- \int_{\lambda{_0}}^\lambda dt \left(\frac{2\gamma(t)}{\beta(t)}\right)\right)T(\lambda , L),\eqno(47)$$
then eq. (46) reduces to
$$\left(f(\lambda) \frac{\partial}{\partial L} + \frac{\partial}{\partial \lambda}\right) T(\lambda , L) = 0\eqno(48)$$
where
$$f(\lambda) = \frac{-1 + \gamma(\lambda) + \beta(\lambda)/\lambda}{\beta(\lambda)} .\eqno(49)$$
Eq. (48) can be solved using separation of variables; if we set $T(\lambda, L) = a(\lambda)b(L)$ then
$$\frac{b^\prime (L)}{b(L)} = \kappa = -\frac{a^\prime (\lambda)}{f(\lambda)a(\lambda)}\eqno(50)$$
where $\kappa$ is a constant. Integration of eq. (50) results in
$$b(L) = Be^{\kappa L}\eqno(51)$$
$$a(\lambda) = A\exp \left(-\kappa \int_{\lambda_0}^\lambda dt f(t)\right)\eqno(52)$$
so that
$$V(\lambda , \phi , \mu) = C\!\!\!\!/ \exp \left(- \int_{\lambda_0}^\lambda dt g(t)\right) e^{\kappa L} \exp \left(-\kappa \int_{\lambda_0}^\lambda dt f(t)\right)\phi^4\eqno(53)$$
where $C\!\!\!\!/ = AB$ and $g(t) = \frac{2\gamma(t)}{\beta(t)}$.

Since $L = \ln \left(\frac{\lambda \phi^2}{2\mu^2}\right)$, all of the dependence on $\phi$ in eq. (53) reduces to just $\phi^{4+2\kappa}$; viz
$$V(\lambda , \phi , \mu) = C\!\!\!\!/ \exp \left[
- \int_{\lambda_0}^\lambda dt \left( g(t) + \kappa f(t)\right)\right] \left(\frac{\lambda \phi^2}{2\mu^2}\right)^\kappa \phi^4.\eqno(54)$$
This supports the possibility in ref. (1) of spontaneous symmetry breaking not occurring in the massless $\phi_4^4$ model when all radiative effects are taken into account.

If in eq. (54) we take $g(t) = \frac{2g}{b}$ and
$f(t) = \frac{-1+gt^2+bt}{bt^2}$, we obtain
$$V(\lambda , \phi , \mu) = C\!\!\!\!/ \exp \left[
- \frac{g}{b} (2 + \kappa)(\lambda - \lambda_0) -  \kappa \left(
\frac{1}{\lambda}- \frac{1}{\lambda_0}\right)\right]\nonumber$$
$$\left(\frac{\lambda}{\lambda_0}\right)^{-\kappa} 
\left(\frac{\lambda\phi^2}{2\mu^2}\right)^\kappa \phi^4.\eqno(55)$$
We now also have the implicit dependence of $V$ on $\mu^2$ governed by the equations
$$\mu^2 \frac{d\lambda}{d\mu^2} = \beta (\lambda) = b\lambda^2\eqno(56)$$
$$\mu^2 \frac{d\phi^2}{d\mu^2} = \gamma_\phi (\lambda)\phi^2 = g\lambda^2
\phi^2 .\eqno(57)$$
Integrating these equations leads to
$$\lambda = \lambda(\mu^2) = \frac{\lambda_1}{1 - \lambda_1 b \ln \left(\frac{\mu^2}{\mu_1^2}\right)}\eqno(58)$$
$$\phi^2 = \phi^2(\mu^2) = \phi_1^2\exp\left(\frac{g\lambda_1}{b} \left(\frac{1}{1-\lambda_1 b\ln\left(\frac{\mu^2}{\mu_1^2}\right)} - 1\right)\right)\eqno(59)$$
where $\lambda_1$ and $\phi_1^2$ are boundary values of $\lambda(\mu^2)$ and $\phi^2(\mu^2)$ at $\mu^2 = \mu_1^2$. Substitution of eqs. (58) and (59) into eq. (55) results in
$$V = C\!\!\!\!/ \exp \left[ \frac{g}{b} (2 + \kappa) (\lambda_0 - \lambda_1) + \kappa \left(\frac{1}{\lambda_0} - \frac{1}{\lambda_1}\right)\right]\left[ \frac{\lambda_0 \phi_1^2}{2\mu_1^2}\right]^\kappa \phi_1^4 .\eqno(60)$$
The renormalization group equation (7) ensures that $V$ is independent of $\mu^2$; this is realized explicitly in eq. (60) when we use the truncated expressions for $\beta$ and $\gamma_\phi$ used in eqs. (56) and (57). If the boundary values $\lambda_0$ and $\lambda_1$ are set equal to each other,  $\kappa$ is set equal to zero, and $C\!\!\!\!/$ is set equal to $\frac{\lambda_1}{4!}$ we obtain simply
$$V = \frac{\lambda_1}{4!} \phi_1^4 .\eqno(61)$$

This disappearance of dependence on $\mu^2$ is similar
to what occurs in the discussion of ``the method of characterisitics'' in ref. [14]. The example used there is the equation
$$x \frac{\partial A(x,y)}{\partial x} + y^2 \frac{\partial A(x,y)}{\partial y} = 0; \eqno(62)$$
a solution of which is
$$A_0 (x,y) = xe^{1/y} .\eqno(63)$$

If ``characteristic functions'' $\overline{x}(t)$ and $\overline{y}(t)$ are defined by 
$$\frac{d\overline{x}(t)}{dt} = \overline{x}(t)\;\;\;\;\; (\overline{x}(0) = x)\eqno(64)$$
$$\frac{d\overline{y}(t)}{dt} = \overline{y}^2(t)\;\;\;\;\; (\overline{y}(0) = y)\eqno(65)$$
then it is evident that $A_0(\overline{x}(t)$, $\overline{y}(t)$) also satisfies eq. (62) and that $\frac{d}{dt} A_0 (\overline{x} (t), \overline{y} (t)) = 0$. Indeed, by eqs. (64) and (65), $\overline{x} (t) = xe^t$ and $\overline{y} (t) = \frac{y}{1-yt}$, and it follows that  
$A_0(\overline{x}(t), \overline{y}(t)) = xe^{1/y}$.  All $t$ dependence has cancelled out. If however we have only considered a perturbative approximation to $A_0(x,y)$ given in eq. (63),
$$A_0^{(1)} (x,y) = x\left(1 + \frac{1}{y}\right) \eqno(66)$$
then $A_0^{(1)} (\overline{x}(t), \overline{y}(t))$ reproduces $A_0(x,y)$ only at the particular value of $t$ given by $t = \frac{1}{y}$. Much the same happens in ref. [10], where a ``boundary function'' for the effective potential is chosen at $L$ loop order to be the computed value of $V$ at that order with $s$ set equal to zero. (All notation is that of ref. [10].). However, this function is not a solution to the full renormalization group equation (7), even if the renormalization group functions are truncated at $L + 1$ loop order; it can be seen that a full non-perturbative solution to the renormalization group equation, even with truncated renormalization group functions (akin to eq. (63), and not eq. (66), providing a solution to eq. (62)) involves portions of all functions $f_{\ell}$ appearing in the sum of eq. (13) of ref. [10].  This accounts for having to set $\overline{s} = 0$ to reproduce the results of ref. [6]; this is much like having to choose $t = \frac{1}{y}$ in order for $A_0^{(1)} (\overline{x}(t), \overline{y}(t))$ to reproduce $A_0(x,y)$.

We are in the process of examining the effective potential for models which scalars possessing a mass at the tree level couple to other fields, such as a gauge Boson.

\section{Acknowledgements}
NSERC provided financial support. FTB would like to thank CNPq and FAPESP for financial support. Alex Buchel had helpful suggestions. Roger Macloud had useful advice. Much of D.G.C. McKeon's work was done while visiting the Perimeter Institute.

\end{document}